**Thickness-dependent Magneto-transport of $Bi_2Se_3/SiO_2$ topological insulator thin films**

Yogesh Kumar[1,2], Prince Sharma[1,2], and V.P.S. Awana[1,2*]

[1]*CSIR-National Physical Laboratory, Dr. K. S. Krishnan Marg, New Delhi-110012, India*
[2]*Academy of Scientific and Innovative Research (AcSIR), Ghaziabad 201002, India*

**Abstract:**

Topological insulators are immensely investigated for their surface states related properties as these materials can be used for various spintronics, quantum computing, and optoelectronics applications. In this perspective, different thicknesses of bismuth selenide thin films are deposited on the 250 nm $SiO_2$ substrate with the help of thermal deposition. The motive of this study is to investigate the surface and bulk-related behaviour with different thicknesses. The deposited films are characterized through GI-XRD (grazing incidence X-ray diffractometer) and Raman spectroscopy, which ensure the impurity less deposition. Further, the transport properties are investigated, which shows thickness dependence of weak anti-localization effect (WAL) in the system and proposed these $Bi_2Se_3/SiO_2$ thin films as a topological Anderson insulator (TAI).

[*]**Corresponding Author**

Dr. V. P. S. Awana:  E-mail: awana@nplindia.org

Ph. +91-11-45609357, Fax-+91-11-45609310

Homepage: awanavps.webs.com

**Introduction:**

Topological insulators (TIs) are now well-known because of their extensive properties, which show the abilities of this kind of quantum materials to be used in different applications in the field of spintronics, optoelectronics, and quantum computing[1–6]. The quests began with the discovery of TIs in the condensed matter community because of its topological-dependent behaviour, which has its conducting surface states and insulating bulk[3, 6, 7]. This duality in the behaviour is protected with the time-reversal symmetry, and it is present due to its intrinsic spin-orbit coupling (SOC)[6–8]. TIs show numerous properties due to their surface states, which can be probed with highly specialized and complex techniques such as angle-resolved photoelectron spectroscopy (ARPES)[9–12]. Moreover, these particular techniques required a single-crystalline sample, which is hard to synthesize. Another way to probe the surface states is to study TIs behaviour at low temperature and low magnetic field because, at this particular condition, the surface states of the TIs are more prominent than the bulk behaviour. Secondly, it can be possible to decrease the thickness of TIs; thus, there is a high chance of dominance of surface states at this condition, as the bulk contribution is decreased through thinning TIs[13–15]. However, there is a reasonable possibility of increasing the defects states with thinner films, which can disturb its intrinsic surface states. Along with these defects states, there is a high probability of finding states which referred to as topological Anderson insulators (TAI)[16–19]. Basically, TAI is a kind of TIs, but with the defects states in it[16–19]. However, the theoretical prediction of TAI is in the literature[16–18], and these TAI can show metal to insulator transition in the TIs.

In this perspective, the thermal evaporation technique is employed to deposit bismuth selenide thin films on the 250 nm $SiO_2$ substrate for probing surface states related transport properties. The deposition of different thicknesses of the thin film is carried out, and its confirmation is characterized through GI-XRD, while the Raman vibration modes and their dependence with thickness are also investigated. Four different thickness films are deposited, which are 5, 10, 15, and 20 nm. The transport properties of all these thin films are investigated with the help of quantum design Physical Property Measurement System (PPMS). It is thought that with decreasing the thickness of the thin films, there is a high probability of an increase in the contribution of surface states. However, the case is not the same here as with a decrease in thickness, and there is an increase in the defects states, which eventually makes 5 nm thin films to be semiconducting, while all other films are metallic in nature. Thus, all these thin films are further investigated for magnetoresistance up to ±12 Tesla in order to see the topological surface states related character. This article is a comprehensive investigation on the thickness-dependent thin films of bismuth selenide, which shows the possibility of TIs being TAI at lower thickness.

**Experimental details:**

The thin films of bismuth selenide are deposited through a thermal evaporation system. The bulk single crystal is used for evaporation where the bulk bismuth selenide is grown under well-optimized heat treatment[20]. The grown crystal is cut down into small pieces, which are put on the molybdenum boat in the thermal evaporation chamber. The $SiO_2$ substrate is also cleaned with IPA (iso-propyl alcohol) and further with nitrogen gas. The cleaned substrate is then placed in the thermal evaporator chamber too. Then, after placing the crystal and substrate, the chamber is sealed, and a vacuum of $10^{-6}$ bar is attained with the help of rotatory and diffusion pumps. The vacuum is achieved in the chamber in approximately 4-5 hrs, and then evaporation of crystal is carried out by increasing the current. The substrates are just placed on the rotator above the molybdenum boat. So at the time of evaporation, the rotator is on with a speed of 45 rpm to get a uniform film.

Moreover, the thickness of the films is observed by thickness monitored of telemark model 851. The deposited films are then annealed at 300º C in the nitrogen environment to have oxidation-free bismuth selenide thin films. Different thicknesses of 5, 10, 15, and 20 nm thin films are deposited using the same process. These deposited thin films are then characterized using GI-XRD and Raman spectroscopy techniques, where the GI-XRD of Panalytical X'pert Pro with Cu-$K_{\alpha 1}$ X-ray radiation which is operating at 40 kV and 40 mA. While the table-top Renishaw is used for investigating Raman vibration modes in the thin films at a laser of 514 nm. Moreover, the transport properties are explored using quantum design Physical Property Measurement System (PPMS) down to 2K, where the chromium-gold pads are deposited using masking in the thermal evaporation technique, and contact on the PPMS puck is made by wire-bounder.

**Results and Discussion:**

The GI-XRD of 5, 10, 15, and 20 nm thin films of bismuth selenide is shown in fig.1 (a). The GI-XRD peaks signify different planes of bismuth selenide, which are *(015), (018), (0012), (1010), (0111), (0015), (1013),* and *(024)*. All these planes are matched with bulk crystals, which ensure the deposition of bismuth selenide in the purest form. However, the full-width half maxima (FWHM) of all these films are found to be dependent on thickness.

Basically, with a decrease in thickness, there is an increase in the FWHM, which signifies the increase of defect states in the system. These thin films are further investigated through Raman spectroscopy to find various Raman vibration modes. There are three Raman modes observed in all these films, which are $A^1_{1g}$, $E^2_g$, and $A^2_{1g}$,[21, 22] as shown in fig.1 (b). These modes are the intrinsic modes of bismuth selenide, which ensure impurity-free deposition. Moreover, there is a Raman peak at 520 cm$^{-1}$ due to SiO$_2$[23], as shown in the inset of fig. 1(b). It is essential to highlight that the FWHM of Raman peak is found to be dependent on thickness in such a way that with decreasing thickness, there is an increase in FWHM. This particular phenomenon supports the earlier argument that with a decrease in the thickness of the films, there is an increase in the defect states in the system. Thus, the GI-XRD and Raman spectroscopy ensure the deposition of pure bismuth selenide, where defects states are induced with decreasing thickness.

The transport properties of all these films are investigated using a quantum design-PPMS system. Fig.2 shows the RT (Resistance vs. temperature curve) curve of all the four thin films down to 2 K. The 10, 15, and 20 nm films show metallic character as their resistance decreases with a decrease in temperature to 20 K, but it starts increasing thereafter down to 2K. The increase in resistance below 20 K signifies some sort of metal to insulator transition. However, the increase is not that large, but it can still be a case of metal to insulator transitions and thus symbolize the concept of TAI [16, 17]. Theoretically, the TAIs are proposed to have metal to insulator transitions due to prominent defect states. Thus, it is possible that the low thickness of thin films and the technique employed to deposit these films have induced defects, which is the reason for such a transition. The 10 and 20 nm thin films almost have the same upward increase from 20 to 2 K, while the 20 nm film shows only a slight push. Thus, it cannot be wrong to say that the concept of TAIs, is only prominent at thinner films, i.e., 10 and 15 nm and above that the bulk contribution of TI comes into the picture, which suppresses the phenomenon of TAIs. Apart from 10, 15, and 20 nm film, the 5 nm film shows an increase in resistance with a lowering of temperature down to 2K. This increase symbolizes the pump-probe spectroscopy study of Glinka et al., which proposed that below 6 nm thickness, there is an opening of Dirac cone and the TIs have a bandgap in their valence and conduction band. The same is the case here, as the increase in resistance ensures the semiconducting nature of 5 nm thin films, which is due to the opening of the Dirac cone. It is vital to highlight one thing here, that there is again a sharp increase in the 5 nm film below 20 K towards 2 K. this sudden increase again supports the concept of TAIs. Thus, the concept of TAIs is only shown in films below 15 nm films as there are more defects states in lower thickness films due to the interface of bismuth selenide and SiO₂, which become the reason for the transition and TIs to be as TAIs. However, at 20 nm film, the bulk states come into the picture and suppress the interface defects.

The RT curve shows the semiconducting to metallic character transition as the thickness of the films go to 5 to 10 nm, while metal to insulator transition as the thinner films goes down to 2K and proposes to behave as TAIs character. These films are then analysed for magnetic field dependence on resistance through RH (resistance vs. magnetic field) curves in order to find the magnetoresistance (MR %) at ±12 Tesla, which is given as

$$MR\% = \frac{\rho_H - \rho_0}{\rho_0} \, X \, 100 \qquad (1)$$

Where, $\rho_H$ and $\rho_0$ are resistivity at H filed and zero filed, respectively. Fig. 3 shows the MR % of all the thin films, where the 5 nm film has ~12% MR at ±12 Tesla, which decreases with increasing thickness. Basically, the 10 and 15 nm film has ~10% MR%, decreasing to ~8% MR% for 20 nm at ±12 Tesla. This decrease symbolizes the increment of bulk contribution in the thin films, while the V-shaped character at the low field in all the thin films shows the presence of the WAL effect, which is an evident property of TIs. Moreover, the V-shaped character has different MR% for different thickness films. As in 5 nm films, the V-shaped character is the reason for ~10% MR, while it is ~ 8% in 10 nm and ~4% in 15 and 20 nm film. These changes in MR% again ensure the bulk character increase with thickness in the bismuth selenide thin films. In 5 and 10 nm films, the MR% is found to be almost saturated above ±4 Tesla, while, in 15 and 20 nm film, it shows a dependence on the magnetic field. It again supports the argument that with an increase in the thickness of thin films, there is an increase in the bulk character of TIs, and there is suppression of surface states related dependence. Further, the HLN (Hikami-Larkin-Nagaoka) is employed to probe the MR response of all these films in order to find various other conduction channels. The HLN equation[24–26] is for change in conductivity of a 2D electron system in the presence of a magnetic field, and it is given by

$$\Delta\sigma(H) = -\frac{\alpha e^2}{\pi h}\left[ln\left(\frac{B_\varphi}{H}\right) - \Psi\left(\frac{1}{2} + \frac{B_\varphi}{H}\right)\right] \qquad (2)$$

Where h is Plank's constant, $\Psi$ is digamma function, e is the electronic charge, H is applied magnetic field and $B_\varphi$ is the characteristic field, and it is given by

$$B_\varphi = \frac{h}{8e\pi L_\varphi^2} \qquad (3)$$

Where $L_\varphi$ is phase coherence length. For theoretically fitting the change in conductivity using HLN equation, the change in magneto-conductivity (MC) is calculated by the difference between conductivity at field H and zero field ($\Delta\sigma(H) = \sigma(H)$- $\sigma(0)$). The HLN parameters have different symbols as $\alpha$ and $L_\varphi$ are used to determine the 2D coherent conducting channel and different types of localization or scattering. Moreover, the value of $\alpha$ can resemble different cases, such as if $\alpha$ is 1, then it is referred to as an orthogonal case, while it is -1/2, then it is a symplectic case. Moreover, if $\alpha$ is 0, it is a unitary case. The $L_\varphi$ predicts the coupling between electron-electron or electron-phonon by inelastic scattering. Fig.4 shows the HLN fitting of MC in all the films, while table. 1 shows the fitting parameters of HLN. The value of $\alpha$ lies in the 0 to -0.5 range at all three temperatures, ensuring the presence of a single conduction channel and, most probably, surface states conduction. However, it contradicts the Glinka et.al theory, as the presence of a single conduction channel and V-shaped MR character ensure the presence of conducting Dirac states, not gapped ones. Thus, the semiconducting resistivity behaviour is due to defects in the thin films.

Moreover, the 10 nm films show a similar range of values as in 5 nm, which confirms the presence of a single conduction channel. While its resistivity is also shown metallic behaviour along with mostly V-shaped behaviour in MR % due to WAL effect confirms the dominance of surface states in 10 nm film. However, on later thickness, i.e., 15 and 20 nm, the $\alpha$ value lies in -0.5 to -1.0, which shows the presence of two conduction channels. The possible conduction channels in 15 and 20 nm films would be surface and bulk conduction channels, respectively. Hence, HLN fitting ensures the presence of a prominent surface conduction channel in 5 and 10 nm film, while the bulk contribution comes into the picture with 15 and 20 nm thin films. However, one thing needs to be highlighted that the HLN model fitting diverges at a high magnetic field of ±1 Tesla. Thus, it is not the best model to probe the intrinsic characteristics of MC at high fields. In order to improvise the HLN fitting, two different parameters are added along with HLN, and the modified equation of theoretical fit of MC is given by

$$\Delta\sigma(H) = -\frac{\alpha e^2}{\pi h}\left[ln\left(\frac{B_\varphi}{H}\right) - \Psi\left(\frac{1}{2} + \frac{B_\varphi}{H}\right)\right] + \beta H^2 + C \qquad (4)$$

Where the first part is again the HLN part, as it best describes the V-shaped contribution due to WAL effect and surface conduction at low field, while at high field, $\beta H^2$ the term comes into the picture. It combines both classical and quantum terms, which correspond to elastic and spin-orbit scattering lengths. At last, the 'c' is added to represent the defect states related contributions. Fig.5 shows the fitting of MC using equation 5, and it is well evident from the fitting that there is no divergence of theoretical fit with experimental results. Table. 2 shows the fitting parameters for $Bi_2Se_3$ thin fims obtained from fitting the HLN+$\beta H^2$+c equation at different temperatures. The value of $\alpha$ lies in the same regime as in the HLN case, while the $\beta$ is found to be decreasing with temperature in all the films, which shows an increase in the bulk contribution. Moreover, the value of $\beta$ is also decreased with an increase in thickness, which also supports the argument of an increase in the bulk contribution with increasing thickness. At last, the value of '$c$' represents the defects state, which shows that there are more defects in 5 nm as compared to the 20 nm thin film of bismuth selenide. In all, this study shows the dependence of the presence of surface and bulk conduction channel with the thickness of the thin films and proposed the possible TAIs in 5, 10, and 15 nm thin films of bismuth selenide, which open up a wide area of applications in the field of spintronics, quantum computing and as a thermistor device.

**Conclusion:**

This article shows the successful deposition of different thicknesses of thin films using the thermal evaporation technique under a well-optimized annealing treatment. The GI-XRD and Raman spectroscopy ensure the impurity-free deposition of all the thin films. While, the transport properties predict the possibility of TAIs in 5, 10, and 15 nm thin films because of metal to insulator-like transitions. Moreover, the MC studies ensure the presence of the WAL effect in all the films along with surface and bulk conduction channel dominance at different thicknesses. The modified HLN fit helps in concluding the defects and bulk contribution in the thin films. This study proposed that the 5, 10, and 15 nm thin films of bismuth selenide over $SiO_2$ are possible experimentally TAIs along with surface states dominance, which can be used as thermistor devices.

**Acknowledgment**

The authors would like to thank Dr. T.N Narayanan and Dr. Rahul Sharma for visit of Mr. Prince Sharma to their institute and helping in thermal deposition and the Raman spectroscopy of the studied thin films. Dr. Mahesh Kumar from CSIR-NPL is acknowledged for his valuable suggestions in regards to annealing of the films. Dr. K.K. Maurya from CSIR-NPL is acknowledged for the GIXRD of the thin films. Also, Prince Sharma and Yogesh Kumar like to thank UGC and CSIR, India, for a research fellowship and AcSIR-NPL for Ph.D. registration.

**Table 1:** HLN analysis up to ±1 Tesla

| Temperature (K) | Thickness (nm) | α | $L_\varphi$(nm) |
|---|---|---|---|
| **2** | 5 | -0.510 | 207.06 |
| | 10 | -0.540 | 185.62 |
| | 15 | -0.620 | 147.99 |
| | 20 | -0.680 | 133.91 |
| **5** | 5 | -0.370 | 188.02 |
| | 10 | -0.463 | 131.87 |
| | 15 | -0.523 | 117.30 |
| | 20 | -0.562 | 110.36 |
| **20** | 5 | -0.318 | 70.82 |
| | 10 | -0.413 | 54.79 |
| | 15 | -0.580 | 51.99 |
| | 20 | -0.754 | 49.32 |

**Table 2:** Extracted parameters for $Bi_2Se_3$ thin films obtained from fitting HLN+$\beta H^2$+c equation at different temperatures.

| Temperature (K) | Thickness (nm) | α | $L_\varphi$(nm) | β | c |
|---|---|---|---|---|---|
| **2** | 5 | -0.347 | 474.55 | 0.00074 | 0.0406 |
| | 10 | -0.443 | 289.01 | 0.00070 | 0.0397 |
| | 15 | -0.564 | 175.39 | -0.00274 | 0.0182 |
| | 20 | -0.680 | 142.86 | -0.00325 | 0.0238 |
| **5** | 5 | -0.257 | 356.05 | 0.00012 | 0.0033 |
| | 10 | -0.387 | 170.10 | 0.00024 | 0.0068 |
| | 15 | -0.481 | 131.89 | -0.00284 | 0.0045 |
| | 20 | -0.678 | 84.11 | -0.00344 | -0.0229 |
| **20** | 5 | -0.208 | 115.29 | -0.00012 | 0.0107 |
| | 10 | -0.323 | 68.52 | -0.00007 | 0.0082 |
| | 15 | -0.462 | 62.54 | -0.00293 | 0.0085 |
| | 20 | -0.761 | 46.03 | -0.00355 | -0.0101 |

**Figure captions:**

**Figure 1(a):** Grazing incidence X-ray diffraction pattern of $Bi_2Se_3$ thin films with various thicknesses.

**Figure 1(b):** The Raman spectrum of $Bi_2Se_3$ thin films at different thicknesses and the inset shows the Raman mode for $SiO_2$.

**Figure 2:** Variation of normalized resistivity ($\rho/\rho_{300K}$) as a function of the temperature of $Bi_2Se_3$ thin films.

**Figure 3:** The variation of magnetoresistance as a function of the transverse magnetic field at different temperatures for $Bi_2Se_3$ thin films with the thickness **(a)** 5nm, **(b)** 10nm, **(c)** 15nm, and **(d)** 20nm.

**Figure 4:** Temperature dependence of magneto-conductivity in a transverse magnetic field is fitted with HLN equation of $Bi_2Se_3$ thin films with the thickness **(a)** 5nm, **(b)** 10nm, **(c)** 15nm, and **(d)** 20nm.

**Figure 5:** Temperature dependence of magneto-conductivity in a transverse magnetic field is fitted with HLN+$\beta H^2$+c equation of $Bi_2Se_3$ thin films with the thickness **(a)** 5nm, **(b)** 10nm, **(c)** 15nm, and **(d)** 20nm.

Figure 1(a):

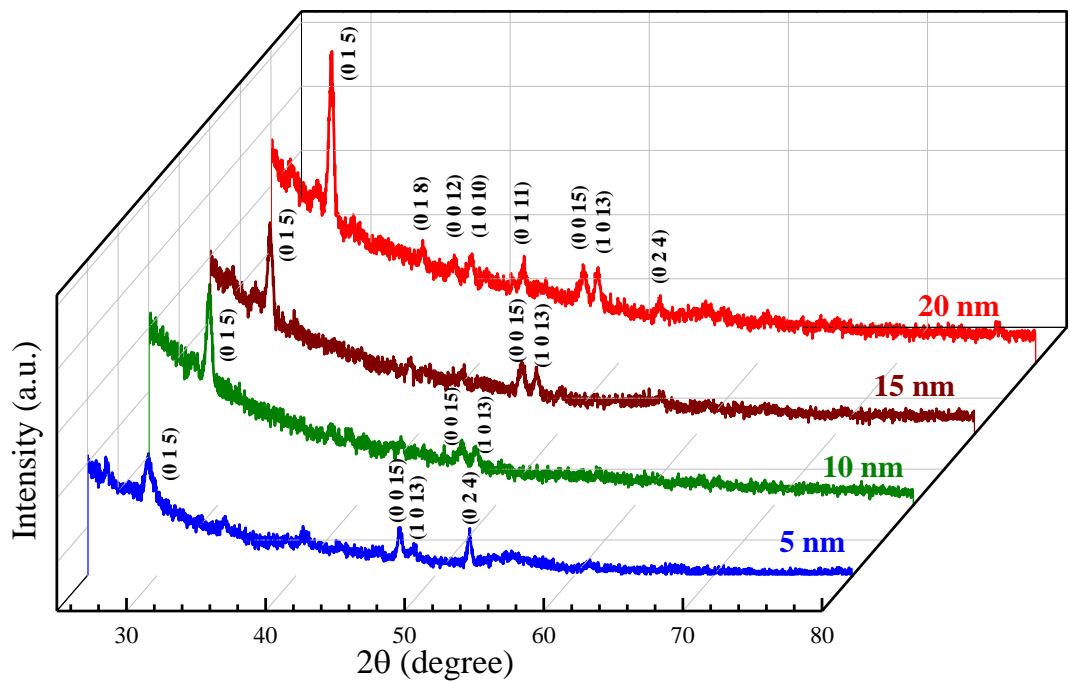

Figure 1(b):

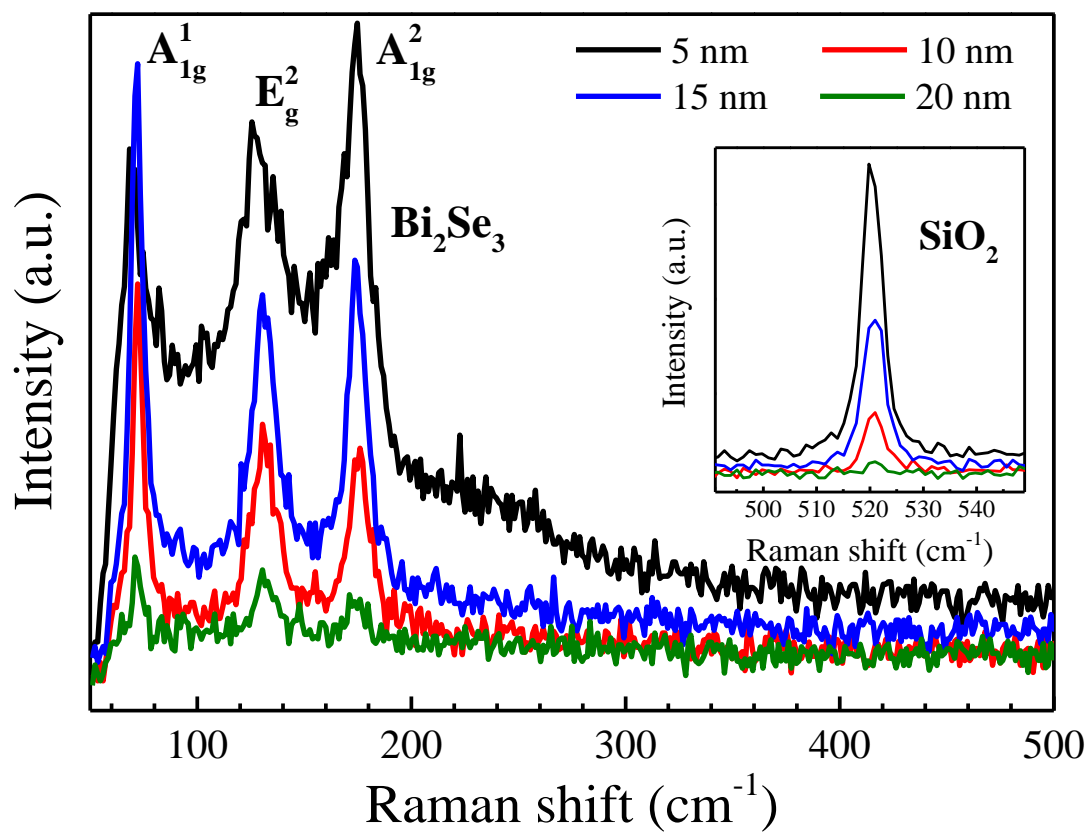

Figure 2:

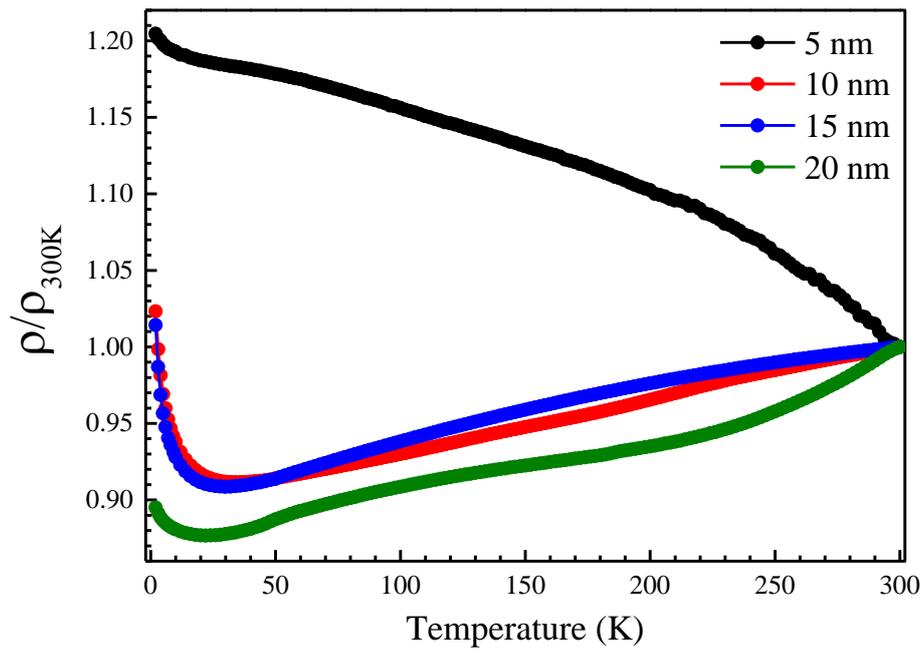

Figure 3:

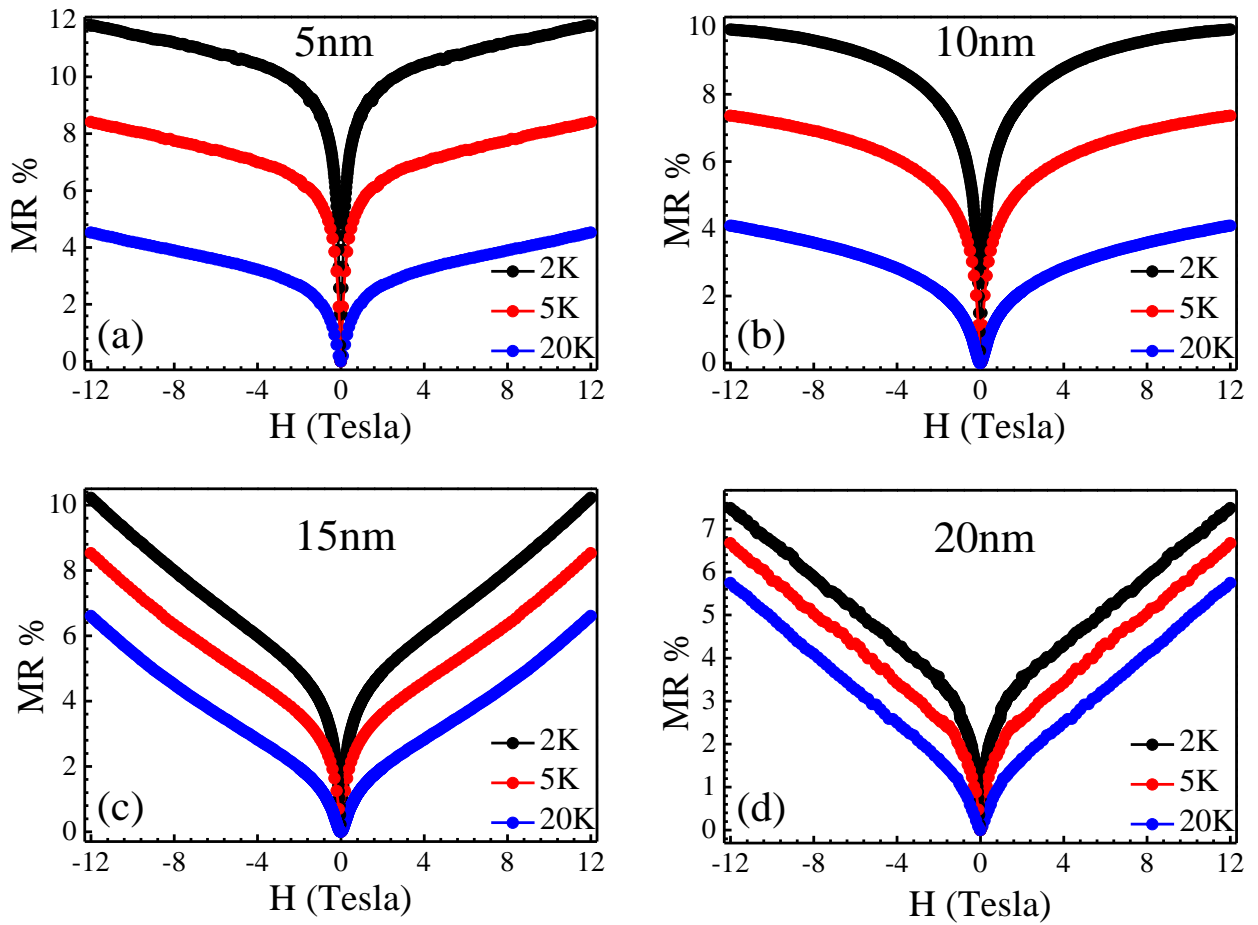

Figure 4:

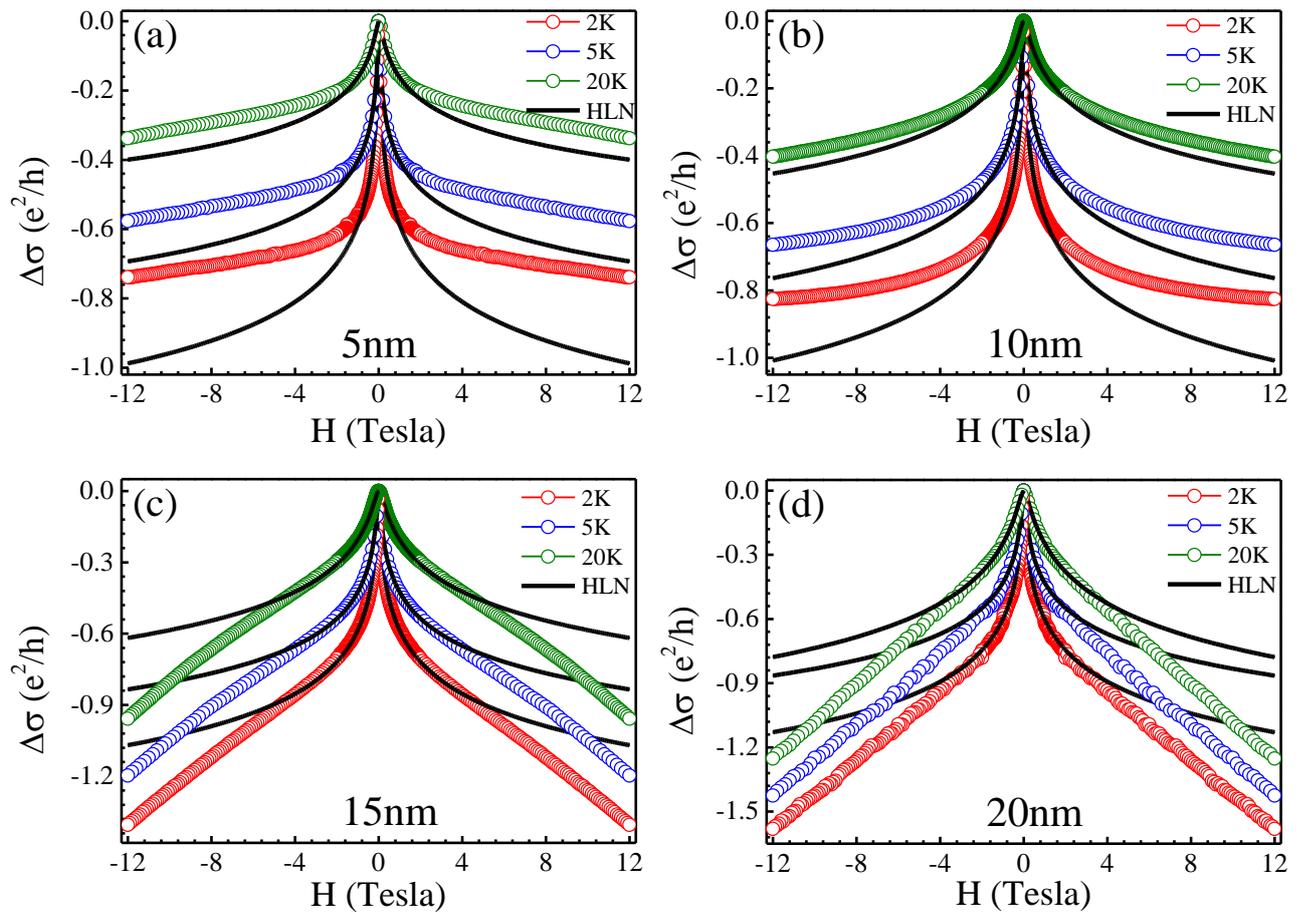

Figure 5:

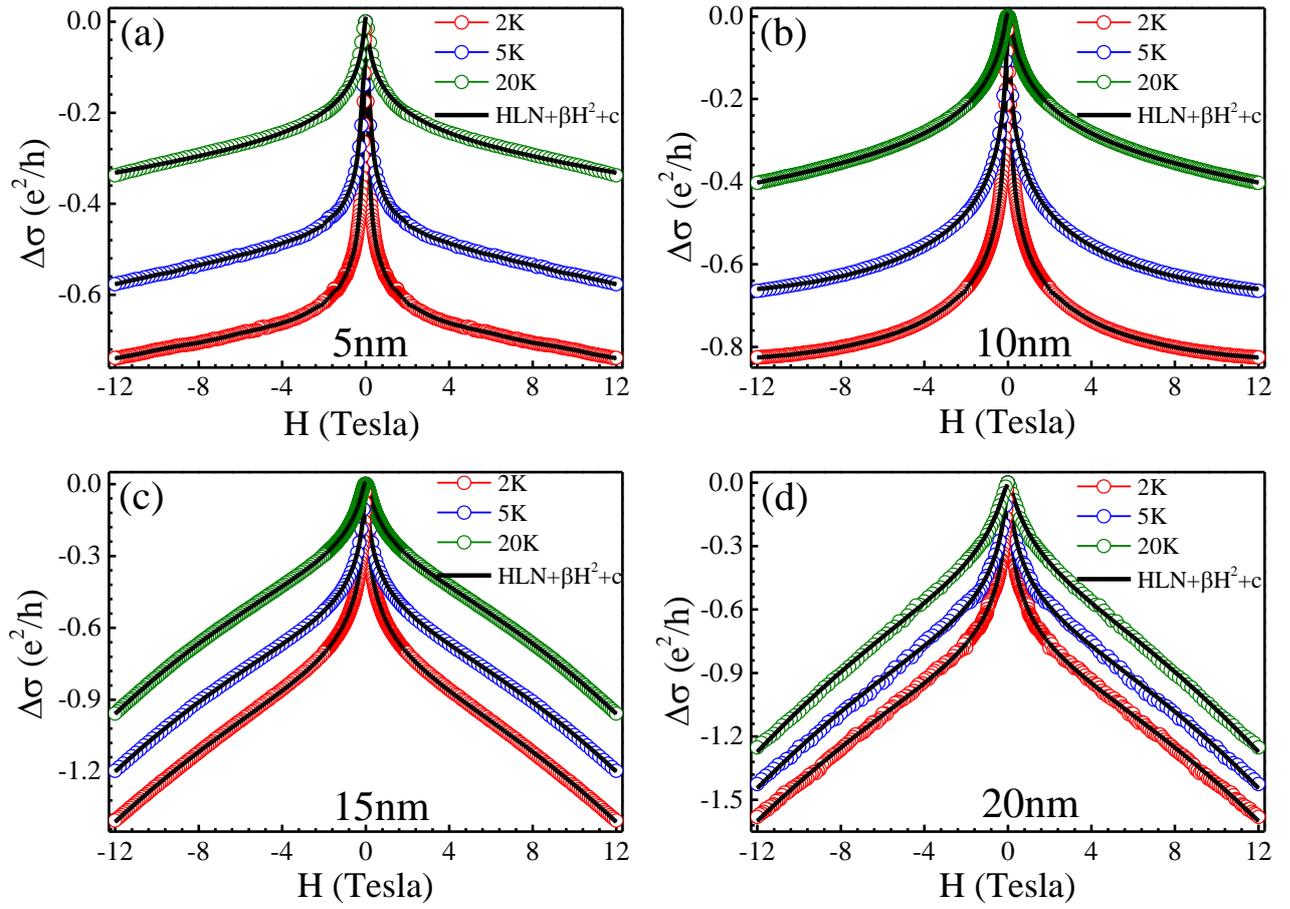